\documentclass{article}
\usepackage[margin=1in]{geometry}
\usepackage{graphicx}
\usepackage{hyperref}
\hypersetup{
    colorlinks=true,
    linkcolor=blue,
    urlcolor=blue,
    citecolor=blue
}

\title{Accelerating Earth Science Discovery via Multi-Agent LLM Systems}
\author{
    Dmitrii Pantiukhin\textsuperscript{1} (\url{https://orcid.org/0000-0002-3427-8188}),\\
    Boris Shapkin\textsuperscript{1} (\url{https://orcid.org/0009-0002-8280-3554}),\\
    Ivan Kuznetsov\textsuperscript{1} (\url{https://orcid.org/0000-0001-5910-8081}),\\
    Antonia Anna Jost\textsuperscript{1} (\url{https://orcid.org/0009-0007-0571-6429}),\\
    Nikolay Koldunov\textsuperscript{1} (\url{https://orcid.org/0000-0002-3365-8146})
}
\date{}

\begin{document}

\maketitle

\begin{center}
    \textsuperscript{1} Alfred Wegener Institute for Polar and Marine Research, Bremerhaven, Germany
\end{center}

\begin{abstract}
    This Perspective explores the transformative potential of Multi-Agent Systems (MAS) powered by Large Language Models (LLMs) in the geosciences. Users of geoscientific data repositories face challenges due to the complexity and diversity of data formats, inconsistent metadata practices, and a considerable number of unprocessed datasets. MAS possesses transformative potential for improving scientists’ interaction with geoscientific data by enabling intelligent data processing, natural language interfaces, and collaborative problem-solving capabilities. We illustrate this approach with "PANGAEA GPT", a specialized MAS pipeline integrated with the diverse PANGAEA database for Earth \& Environmental Science, demonstrating how MAS-driven workflows can effectively manage complex datasets and accelerate scientific discovery. We discuss how MAS can address current data challenges in geosciences, highlight advancements in other scientific fields, and propose future directions for integrating MAS into geoscientific data processing pipelines. In this Perspective, we show how MAS can fundamentally improve data accessibility, promote cross-disciplinary collaboration, and accelerate geoscientific discoveries.
\end{abstract}

\section{Introduction}
Geoscience data archives, which serve as curated digital infrastructures for the systematic storage and distribution of Earth and environmental datasets, have grown to enormous scales, with large repositories such as PANGAEA, NASA’s Earth Observing System Data and Information System (EOSDIS), NOAA’s National Centers for Environmental Information (NCEI), and the Copernicus Climate Data Store (C3S) collectively hosting millions of heterogeneous datasets and several petabytes of data \cite{kobler1995, felden2023}. For example, PANGAEA alone contains more than 400,000 datasets derived from a variety of observational platforms ranging from shipboard expeditions and sediment cores to global climate model outputs \cite{felden2023}. Despite this scale, data citation remains low, with over 93\% of datasets being uncited \cite{robinson2016}. At the core of this issue with data reuse are inadequate metadata standards, non-uniform data formats, and incomplete documentation \cite{gil2016}. As a result, countless datasets remain underutilized or completely absent from scientific publications, resulting in missed opportunities for scientific discovery.

These challenges of managing and utilizing complex, heterogeneous datasets extend far beyond geosciences, impacting a wide array of scientific disciplines, where similar issues with data formats and metadata persist \cite{zhang2015, schnase2016, pal2020}. Large Language Models (LLMs), with their ability to parse unstructured data and reason across domains, provide a robust foundation for addressing these challenges \cite{bauer2024}. LLMs have become powerful tools for information retrieval and analysis in various scientific fields \cite{guo2024}. Recent GPT models use advanced techniques like reinforcement learning and chain-of-thought reasoning \cite{wei2022}. They excel at complex scientific problems, even surpassing PhD-level experts on tough benchmarks \cite{rein2023, guo2025}. These models can now perform multi-step reasoning, breaking down complex problems into manageable subtasks and synthesizing information from vast knowledge bases. Moreover, they can operate with tools and execute commands, expanding their problem-solving capabilities \cite{paranjape2023}. These advanced capabilities suggest that LLM-driven approaches hold tremendous promise for geoscience data management.

A further evolution of LLM is expected in an agent-based mode, where models function as autonomous agents capable of performing tasks without constant human guidance, relying on tools, databases, internal memory and other resources \cite{guo2024}. In this context, an agent is an LLM-based system that can perceive the environment, reason about the information it receives, and take actions to achieve specific goals. Such agents are often used collaboratively in a divide-and-conquer approach, deploying multiple specialised agents that can solve complex analytical problems as a group \cite{qian2023}. This is particularly relevant in geosciences, where the diversity of data formats and the need for domain-specific expertise are significant challenges. By working together, these agents can efficiently handle heterogeneous datasets, bridge gaps between different branches of geoscience, and provide researchers with more holistic insights.

Single-agent and chat-completion approaches have already shown practical benefits in geosciences. Retrieval-augmented techniques (RAG) \cite{lewis2020} which enable language models to dynamically access and incorporate information from external knowledge bases, have particularly enhanced domain-specific accuracy in climate science. For example, RAG approaches with curated climate reports have improved domain-specific QA tasks derived from Intergovernmental Panel on Climate Change (IPCC) documents \cite{vaghefi2023}. Similarly, efforts such as “ClimSight” now provide climate projection information to non-specialist users by integrating LLMs with climate model data \cite{koldunov2024}. Recent advances in RAG techniques have moved beyond simple document retrieval, incorporating multi-level retrieval mechanisms and knowledge graphs to enhance contextual understanding \cite{edge2024}. These developments in retrieval-based systems are rapidly evolving and promise more sophisticated and accurate interactions with domain-specific knowledge bases.

Furthermore, several groups have explored ways to adapt general-purpose LLMs to the geosciences by further training them on large domains of geoscientific data. Notable work includes K2 \cite{deng2024} and GEOGALACTICA \cite{lin2023}, which introduced new benchmarks and datasets for geoscience-specific tuning, as well as OceanGPT \cite{bi2023}, which targets oceanographic tasks, and ClimateGPT \cite{thulke2024}, which is fine-tuned on climate-related data.

Tool integration is a central technical feature of LLM based agents \cite{guo2024}. Instead of relying solely on an LLM’s internal weights, MAS agents leverage tool wrappers, dynamic function calls, and API endpoints to execute domain-specific operations \cite{gim2024}. In the geosciences, this integration has enabled the creation of specialized systems tailored to complex data and analysis needs. For example, Chen et al. proposed “GeoAgent,” a specialized LLM-based framework for geospatial data analysis that integrates a code interpreter, static analysis, and RAG \cite{chen2024}. Another common use of single-agent systems with tool integration is in the application of search capabilities \cite{sun2023}. For geosciences, an illustrative example is LLM-Find \cite{ning2025}, which focuses on geospatial data extraction by providing LLMs with iterative debugging capabilities to retrieve spatial datasets (e.g., OpenStreetMap, weather APIs).

Although these projects have advanced LLM fine-tuning and the use of RAG and tools for geoscience challenges, no multi-agent frameworks have yet been developed that are tightly integrated with underlying geoscience databases.

\section{Emergence of the MAS}
Multi-agent LLM systems represent a major shift from single, isolated AI assistants to distributed architectures of specialized, task-oriented agents \cite{guo2024}. In such systems, each agent operates under predefined sets of instructions, and is equipped with domain-specific reasoning modules, customized knowledge databases, and direct interfaces to external tools and computational sandboxes. Advanced coordination strategies, frequently organized in hierarchical or graph-based flows, enable these agents to exchange intermediate results, negotiate optimal workflows, and iteratively refine partial outputs through chain-of-thought reasoning and reflection \cite{agashe2024, pan2024}. While MAS have shown impressive results in other domains, including collaborative code-generation in software engineering \cite{qian2023, hong2023}, coordinated planning in multi-robot systems \cite{mandi2023}, modeling of complex societal interactions \cite{park2023}, and strategic reasoning in game simulation \cite{wang2023}, no integrated MAS solution has yet been applied to geoscientific data archives.

Nevertheless, initial MAS prototypes for geoscience-related tasks have begun to emerge. For example, ShapefileGPT \cite{lin2024} demonstrated a two-agent LLM framework for automating GIS shapefile processing, where a planner agent delegates spatial subtasks to a worker agent via a specialized function library. Another project, GeoLLM-Squad \cite{lee2025}, introduced a multi-agent paradigm to remote sensing workflows by separating an orchestration agent from multiple domain-specific sub-agents, using open-source frameworks such as AutoGen to integrate modular API toolchains, interactive map UIs, intent-based tool selection, and workflow storage.

In practice, MAS architectures grounded in LLMs can span a wide spectrum of organizational structures, ranging from a single coordinating supervisor to fully autonomous “swarm” networks that collaborate without centralized control \cite{guo2024}. Centralized systems rely on a top-level planner or “supervisor” agent \cite{qian2023} that breaks down tasks, delegates them to specialized sub-agents (e.g., retrieval, analytical, data transformation, validation), and then synthesizes final outputs. This approach, exemplified by hierarchical frameworks such as HuggingGPT \cite{shen2024}, ensures a clear command-and-control mechanism, simplifies quality checks, and promotes consistent workflow management.

By contrast, decentralized models draw inspiration from social systems and swarm intelligence, letting each LLM agent operate more independently with local memory and goals, leading to emergent behaviors and robust parallelization \cite{huang2024}. Hybrid approaches combine both strategies - for instance, dynamic orchestration via a transient “lead” agent while other agents freely negotiate tasks or refine each other’s outputs, mirroring human team dynamics \cite{huang2024}.

One of the advantages of MAS is the ability for multiple agents to simultaneously use specialized tools, each solving different components of a complex problem. Examples might include invoking geospatial libraries such as Geospatial Data Abstraction Library (GDAL) for coordinate transformation, using NetCDF \cite{rew1990} or xarray \cite{hoyer2017} to parse and aggregate spatiotemporal data cubes, and running specialized Python or R scripts for statistical analysis. Agents can perform iterative refinement steps \cite{madaan2024}, re-checking results against data integrity constraints, filtering outliers using robust statistical thresholds, or querying uncertainty quantification modules that assess the credibility of results. This tool ecosystem allows MAS to move beyond static text generation, facilitating a closed-loop interaction model where data retrieval, pre-processing, quality control, analysis, and visualization occur under the guidance of autonomous, domain-aware agents. Reflection and self-critique loops can be implemented by designating a “validator” agent that routinely inspects outputs for internal consistency, methodological rigor, and adherence to community standards. Such approaches use iterative improvement pipelines that break down instructions into smaller steps, critique initial results, and apply further improvements \cite{ferraz2024}.

The agent ecosystem is heterogeneous and includes various specialized agents-such as retrieval agents, analytical agents, data conversion agents, and reporting agents-that work together to accomplish various data management tasks \cite{guo2024}. Retrieval agents incorporate retrieval-augmented generation (RAG) techniques, coupling embeddings from domain-specialized language models with vector databases that index geoscientific literature, vocabularies, and reference datasets \cite{lewis2020}. Analytical agents may run topological anomaly detection on bathymetric grids, apply wavelet transforms to paleoclimate proxies, or compute ensemble mean biases in Coupled Model Intercomparison Project (CMIP) - class climate model runs \cite{eyring2016}. Transformation agents handle unit conversions, project datasets onto common spatial grids, or standardize attribute names. Reporting agents synthesize results into structured outputs, annotate data lineage, and cite relevant publications. RAG-based knowledge infrastructures leverage curated metadata schema and persistent semantic stores that retain cross-session memory, allowing MAS to gradually refine a hypothesis or revisit previously unexplained anomalies. Iterative reasoning loops that incorporate domain feedback can detect subtle teleconnections in ocean-atmosphere systems, illuminate previously unrecognized correlations in coastal sedimentary records, or integrate high-resolution satellite measurements with legacy chemical tracers to map the evolution of marine biogeochemical cycles.

In the context of applying MAS to geoscience tasks, MAS can mimic the dynamics of interdisciplinary research teams, where specialists contribute their expertise, as has been done in software engineering \cite{qian2023}. This synergy is essential for tackling challenges in Earth sciences, from predicting the response of ocean circulation to future warming scenarios to detecting subtle geologic signals of hazard precursors in tectonically active regions. The integration of specialized agents, robust tool usage layers, RAG-based semantic indexing, and adaptive architectural principles would establish MAS as advanced computational platforms for geoscientific discoveries.

\section{PANGAEA GPT: MAS architecture for geoscientific data discovery}
\begin{figure}[h]
    \centering
    \includegraphics[width=0.95\textwidth]{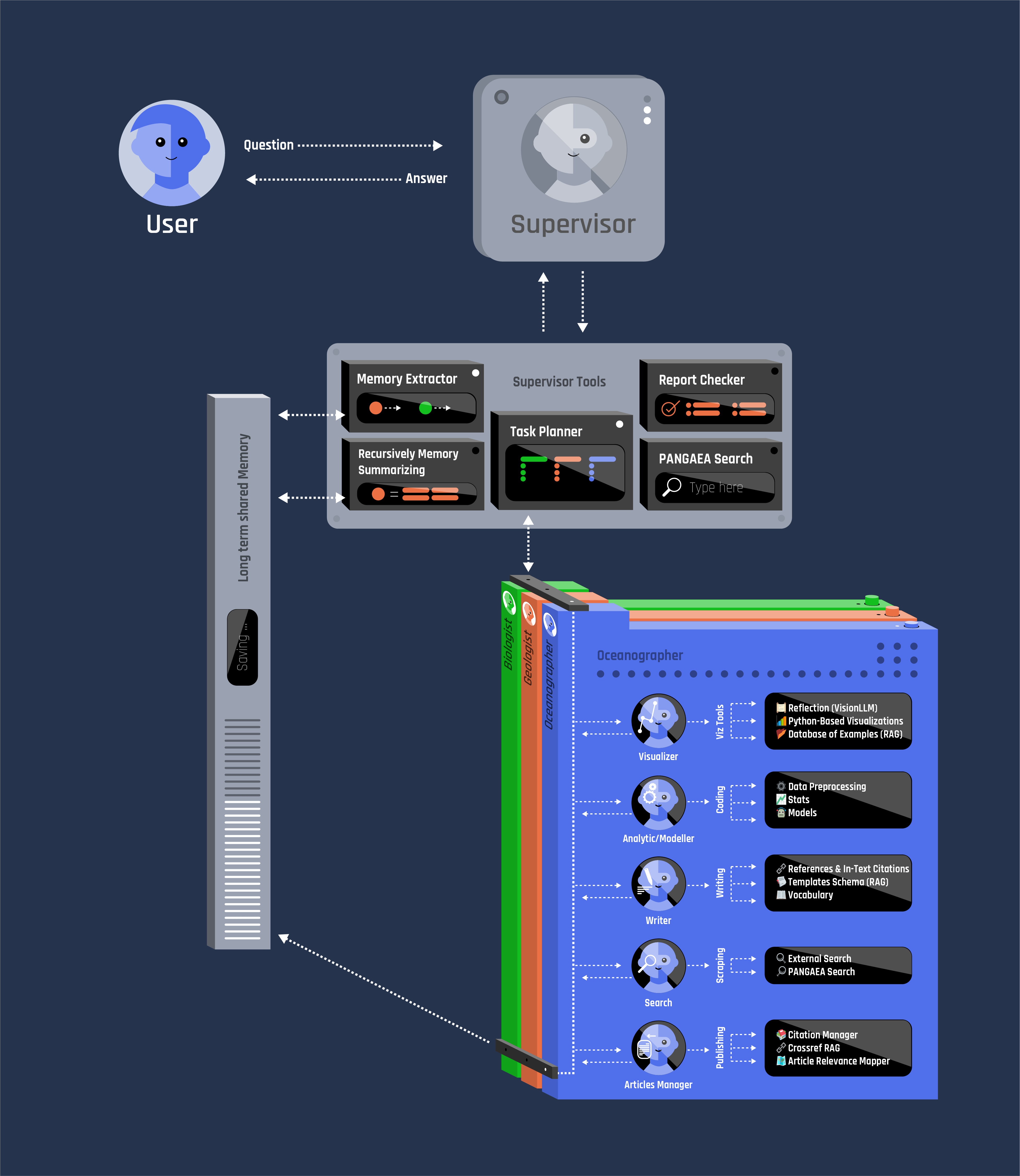}
    \caption{Conceptual framework Multi-agent system (MAS) for geo-scientific data discovery.}
    \label{fig:mas}
\end{figure}

To illustrate how the guiding principles described earlier can be put into practice, we propose a multi-agent system (MAS) architecture specifically designed for geoscience data management - focusing on large and diverse repositories such as PANGAEA \cite{felden2023}. Based on our experience developing PANGAEA GPT - an open-source, LLM-driven multi-agent framework (publicly available at \url{github.com/dmpantiu/pangaeaGPT} and testable at \url{huggingface.co/spaces/CliDyn/pangaeagpt}) we illustrate how a centralized orchestration approach, where a supervisor agent directs domain-specific sub-agents (e.g., in oceanography, biology and geology), can be effectively implemented in geoscience contexts. This modular architecture allows the supervisor to spawn sub-agents on demand, adapting the system’s capabilities to the unique demands of each query. By referencing PANGAEA as a prime example of a heterogeneous database with unconventional formats and challenging metadata, we demonstrate how such a system can handle complex data workflows and provide robust reporting (Fig.~\ref{fig:mas}).

We chose a centralized orchestration approach, where the supervisor agent serves as the command-and-control node for the entire pipeline, handling sub-task delegation, resource allocation, and consolidation of final results among specialized agents (Fig.~\ref{fig:mas}). The system supports the dynamic creation of specialized agents based on the tasks assigned to it. Upon receiving user requests, the supervisor agent constructs agent subgraphs tailored to specific subdomains - oceanography, geology, climatology, ecology, or others - depending on the nature of the query. Each agent operates with localized memory buffers for context retention, set of tools and retrieval-augmented generation (RAG) capabilities that draw upon curated knowledge sources. Such a system is designed to efficiently search through diverse data collections, perform contextual analyses, produce high-quality visualizations, and ultimately generate comprehensive documentation.

A key foundation of this architecture is the use of external tools to address domain-specific analysis needs (Fig.~\ref{fig:mas}). Upon deployment, each agent is provided with a dedicated "sandbox" containing domain-specific software, pre-installed packages, and the necessary data (e.g. bathymetry, seafloor topography maps, multispectral satellite imagery, ocean color data, paleoclimate proxy records, atmospheric reanalysis fields, and etc.). In addition, the agent's operational environment includes a RAG-accessible repository of sample visualizations, statistical analyses, and domain-specific workflows. These features enhance the agent's accuracy in answering user queries, reduce hallucinations by providing reliable domain references, and shorten reflection cycles by enabling rapid retrieval of reliable examples. In our PANGAEA GPT implementation, agents can also critically evaluate their outputs by invoking specialized reflection modules (both for statistical validation and visual question answering). These modules provide constructive feedback - both on statistical results and on visual representations (via visual question answering) and guide the agent through iterative refinements until the final outputs meet the required quality standards.

In addition to assigning tasks, the supervisor agent handles memory and manages information flow across the system (Fig.~\ref{fig:mas}). Based on our experience running PANGAEA GPT, a multi-tier memory approach (storing short-term data in active memory and long-term data in a searchable database) was particularly effective for long-running sessions \cite{liu2024}. Each agent runs locally and, after finishing its cycle, sends outputs back to the supervisor. To avoid overloading, the supervisor monitors resource usage, summarizes logs into short blocks, and then moves them into the long-term RAG database. Short-term context remains in the model’s direct context window, while extended data or partial results are stored in a vector database, retrievable on demand. This setup - short-term context paired with a stable long-term store-supports multi-step exploration without sacrificing critical details, and lowers computational costs during elaborate sessions.

A major challenge we encountered while deploying PANGAEA GPT is verifying the correctness and relevance of multi-agent LLM outputs in the face of highly varied geoscientific data. We believe the lessons learned here can guide future multi-agent LLM implementations facing similar constraints, as unlike software engineering, which typically uses standardized test suites or automated Quality Assurance (QA) workflows \cite{jimenez2023}, Earth science has only a few domain-specific benchmarks that accommodate specialized terminologies and heterogeneous data \cite{bi2023, zhang2024}. Another significant issue is the lack of any “imaging benchmark” that covers the range of visualization practices, which is further complicated by the fact that different programming languages are commonly used - ecologists or biologists often rely on R for plots \cite{gao2025}, while oceanographers tend to prefer Python or Matlab. This diversity translates into an equally broad spectrum of plot types - from distribution maps and cross-sectional charts to correlation matrices - each governed by domain-specific conventions that generic validators rarely catch.

Our experience suggests that the solution for these domain-specific checks is specialized validation agents with their own checklists. For instance, they can confirm whether depth axes are reversed correctly or whether unit scales match geoscientific norms. In PANGAEA GPT, these validation agents currently cross-check outputs against standardized vocabularies (e.g., SeaDataNet) and domain-specific norms, such as ensuring reversed depth axes in oceanographic plots, with ongoing efforts to expand their scope to handle diverse visualization practices and statistical rigor across subdomains. By flagging suspicious metadata entries or unusual variable usage, these agents act as quality-control gates at each major step (data retrieval, analysis, and final visualization). While a universal benchmark for LLM validation remains an important goal, our work indicates that domain-focused modules are essential - particularly for detailed imaging tasks.

\section{Outlook}
Looking forward, the integration of MAS systems into geoscientific research opens up entirely new opportunities for revitalizing previously underutilized historical data as well as more recently generated geoscientific data sources. Autonomous agent networks could systematically explore large repositories, identifying and summarizing historical datasets that have remained underutilized. By combining database-search agents with domain-specific expert analytic agents, this approach can help re-explore entire historical databases and interpret understudied collections. Ultimately, such a system may facilitate renewed engagement with valuable historical data and potentially give rise to new discoveries.

Another potentially promising direction is the use of MAS systems to assist in expedition planning. In the domain of shipping and commercial sectors, LLMs are already being proposed for planning to optimize routes and enhance safety \cite{pei2024}. A potential MAS structure could be envisioned in which one agent first requests historical expedition data (taken directly from PANGAEA or other repositories), another agent checks current satellite products and climate forecasts, and yet another agent integrates predicted weather conditions and ocean currents. Together, these specialized agents would generate individualized expedition plans that optimize time spent at stations, for example, by coordinating dive schedules and sampling activities based on dynamic environmental factors. Such adaptability can streamline logistics and mitigate risks, particularly in remote or high-latitude regions prone to rapid weather changes, ultimately improving both the cost-effectiveness and safety of scientific expeditions. Additionally, the increasing use of Autonomous Underwater Vehicles (AUVs) in modern observatories \cite{wynn2014} makes MAS particularly attractive for operating such fleets during expeditions. These systems could control the AUVs, manage real-time data collection and transmission to repositories like PANGAEA, and use the analyzed data to dynamically re-optimize the AUVs' routes, thereby enhancing the overall efficiency and adaptability of scientific missions.

A more radical idea would be to create a self-sufficient structure of autonomous, wandering chains of agents. One of the most far-reaching goals for MAS in geosciences is the formation of "wandering," self-organizing systems of agents that continuously explore repositories, generating and testing new hypotheses without direct human guidance. These autonomous agent networks could continuously explore the scientific data landscape within repositories, suggesting new directions for research. Relying on unsupervised anomaly detection routines, they would be able, for example, to detect unexpected signals in real-time global seismic data or satellite ocean color imagery, thereby revealing potential new phenomena or hazard precursors. At the same time, a supervisor agent can spawn subordinate agents to propose mechanistic explanations for each anomaly, linking them to known processes. If the system would find plausible but unconfirmed signals, such as a region of unusual phytoplankton bloom, it could trigger additional analyses or domain-expert agents to investigate further, eventually communicating the summarized results to human scientists for more in-depth validation. Over time, this feedback loop could significantly shorten the time between discovery and research action, accelerating environmental insights that might otherwise remain buried in massive data stores. Such a self-governing swarm of agents could directly serve both experts and the general public, democratizing access to research data and broadening the scope of communication.

\section{Acknowledgments}
We thank the Helmholtz Association and the Federal Ministry of Education and Research (BMBF) for supporting the DataHub Initiative of the Research Field Earth and Environment, that supported this study. This work was also supported by the European Union’s Destination Earth Initiative and relates to tasks entrusted by the European Union to the European Centre for Medium-Range Weather Forecasts implementing part of this Initiative with funding by the European Union. This work is also supported by projects S1: Diagnosis and Metrics in Climate Models of the Collaborative Research Centre TRR 181 “Energy Transfer in Atmosphere and Ocean”, funded by the Deutsche Forschungsgemeinschaft (DFG, German Research Foundation, project no. 274762653). Views and opinions expressed are those of the authors only and do not necessarily reflect those of the European Union or the European Climate Infrastructure and Environment Executive Agency (CINEA). Neither the European Union nor the granting authority can be held responsible for them. We thank Thomas Jung for helpful discussions regarding this study. We also thank Jan Wehner for his assistance with figure design.


\begin{thebibliography}{99}
\bibitem{kobler1995}
Kobler, B., Berbert, J., Caulk, P. \& Hariharan, P. C. Architecture and design of storage and data management for the NASA Earth Observing System Data and Information System (EOSDIS). \textit{Proceedings of IEEE 14th Symposium on Mass Storage Systems} 65--76 (1995). \url{https://doi.org/10.1109/MASS.1995.528217}

\bibitem{felden2023}
Felden, J., Möller, L., Schindler, U. \textit{et al.} PANGAEA - Data Publisher for Earth \& Environmental Science. \textit{Scientific Data} 10, 347 (2023). \url{https://doi.org/10.1038/s41597-023-02269-x}

\bibitem{robinson2016}
Robinson-García, N., Jiménez-Contreras, E. \& Torres-Salinas, D. Analyzing data citation practices using the Data Citation Index. \textit{Journal of the Association for Information Science and Technology} 67, 2964--2975 (2016). \url{https://doi.org/10.1002/asi.23529}

\bibitem{gil2016}
Gil, Y. \textit{et al.} Toward the Geoscience Paper of the Future: Best practices for documenting and sharing research from data to software to provenance. \textit{Earth and Space Science} 3, 388--415 (2016). \url{https://doi.org/10.1002/2015EA000136}

\bibitem{zhang2015}
Zhang, Y. \& Zhao, Y. Astronomy in the big data era. \textit{Data Science Journal} 14, 11--11 (2015).

\bibitem{schnase2016}
Schnase, J. L., Lee, T. J., Mattmann, C. A., Lynnes, C. S., Cinquini, L., Ramirez, P. M. \& Carriere, L. Big data challenges in climate science: Improving the next-generation cyberinfrastructure. \textit{IEEE Geoscience and Remote Sensing Magazine} 4(3), 10--22 (2016).

\bibitem{pal2020}
Pal, S., Mondal, S., Das, G., Khatua, S. \& Ghosh, Z. Big data in biology: the hope and present-day challenges in it. \textit{Gene Reports} 21, 100869 (2020).

\bibitem{bauer2024}
Bauer, P., Hoefler, T., Stevens, B. \textit{et al.} Digital twins of Earth and the computing challenge of human interaction. \textit{Nature Computational Science} 4, 154--157 (2024). \url{https://doi.org/10.1038/s43588-024-00599-3}

\bibitem{guo2024}
Guo, T., Chen, X., Wang, Y., Chang, R., Pei, S., Chawla, N. V. \& Zhang, X. Large language model based multi-agents: A survey of progress and challenges. Preprint at \url{https://arXiv.org/abs/2402.01680} (2024).

\bibitem{wei2022}
Wei, J., Wang, X., Schuurmans, D., Bosma, M., Xia, F., Chi, E., ... \& Zhou, D. Chain-of-thought prompting elicits reasoning in large language models. \textit{Advances in Neural Information Processing Systems} 35, 24824--24837 (2022).

\bibitem{rein2023}
Rein, D., Hou, B. L., Stickland, A. C., Petty, J., Pang, R. Y., Dirani, J., ... \& Bowman, S. R. Gpqa: A graduate-level google-proof q\&a benchmark. Preprint at \url{https://arXiv.org/abs/2311.12022} (2023).

\bibitem{guo2025}
Guo, D., Yang, D., Zhang, H., Song, J., Zhang, R., Xu, R., ... \& He, Y. Deepseek-r1: Incentivizing reasoning capability in llms via reinforcement learning. Preprint at \url{https://arXiv.org/abs/2501.12948} (2025).

\bibitem{paranjape2023}
Paranjape, B., Lundberg, S., Singh, S., Hajishirzi, H., Zettlemoyer, L. \& Ribeiro, M. T. Art: Automatic multi-step reasoning and tool-use for large language models. Preprint at \url{https://arXiv.org/abs/2303.09014} (2023).

\bibitem{qian2023}
Qian, C., Cong, X., Yang, C., Chen, W., Su, Y., Xu, J., ... \& Sun, M. Communicative agents for software development. Preprint at \url{https://arXiv.org/abs/2307.07924} (2023).

\bibitem{lewis2020}
Lewis, P., Perez, E., Piktus, A., Petroni, F., Karpukhin, V., Goyal, N., Küttler, H., Lewis, M., Yih, W.-t., Rocktäschel, T., Riedel, S. \& Kiela, D. Retrieval-Augmented Generation for Knowledge-Intensive NLP Tasks. Preprint at \url{https://arXiv.org/abs/2005.11401} (2020).

\bibitem{vaghefi2023}
Vaghefi, S. A., Stammbach, D., Muccione, V. \textit{et al.} ChatClimate: Grounding conversational AI in climate science. \textit{Communications Earth \& Environment} 4, 480 (2023). \url{https://doi.org/10.1038/s43247-023-01084-x}

\bibitem{koldunov2024}
Koldunov, N. \& Jung, T. Local climate services for all, courtesy of large language models. \textit{Communications Earth \& Environment} 5, 13 (2024). \url{https://doi.org/10.1038/s43247-023-01199-1}

\bibitem{edge2024}
Edge, D., Trinh, H., Cheng, N., Bradley, J., Chao, A., Mody, A., ... \& Larson, J. From local to global: A graph rag approach to query-focused summarization. Preprint at \url{https://arXiv.org/abs/2404.16130} (2024).

\bibitem{deng2024}
Deng, C., Zhang, T., He, Z., Chen, Q., Shi, Y., Xu, Y., ... \& He, J. K2: A foundation language model for geoscience knowledge understanding and utilization. \textit{Proceedings of the 17th ACM International Conference on Web Search and Data Mining} 161--170 (2024).

\bibitem{lin2023}
Lin, Z., Deng, C., Zhou, L., Zhang, T., Xu, Y., Xu, Y., ... \& Zhou, C. Geogalactica: A scientific large language model in geoscience. Preprint at \url{https://arXiv.org/abs/2401.00434} (2023).

\bibitem{bi2023}
Bi, Z., Zhang, N., Xue, Y., Ou, Y., Ji, D., Zheng, G. \& Chen, H. Oceangpt: A large language model for ocean science tasks. Preprint at \url{https://arXiv.org/abs/2310.02031} (2023).

\bibitem{thulke2024}
Thulke, D., Gao, Y., Pelser, P., Brune, R., Jalota, R., Fok, F., ... \& Goldstein, H. Climategpt: Towards ai synthesizing interdisciplinary research on climate change. Preprint at \url{https://arXiv.org/abs/2401.09646} (2024).

\bibitem{gim2024}
Gim, I., Lee, S. S. \& Zhong, L. Asynchronous LLM Function Calling. Preprint at \url{https://arXiv.org/abs/2412.07017} (2024).

\bibitem{chen2024}
Chen, Y., Wang, W., Lobry, S. \& Kurtz, C. An llm agent for automatic geospatial data analysis. Preprint at \url{https://arXiv.org/abs/2410.18792} (2024).

\bibitem{sun2023}
Sun, W., Yan, L., Ma, X., Wang, S., Ren, P., Chen, Z., ... \& Ren, Z. Is ChatGPT good at search? investigating large language models as re-ranking agents. Preprint at \url{https://arXiv.org/abs/2304.09542} (2023).

\bibitem{ning2025}
Ning, H., Li, Z., Akinboyewa, T. \& Lessani, M. N. An autonomous GIS agent framework for geospatial data retrieval. \textit{International Journal of Digital Earth} 18, 2458688 (2025).

\bibitem{agashe2024}
Agashe, S., Fan, Y., Reyna, A. \& Wang, X. E. Llm-coordination: evaluating and analyzing multi-agent coordination abilities in large language models. Preprint at \url{https://arXiv.org/abs/2310.03903} (2024).

\bibitem{pan2024}
Pan, B., Lu, R., Wang, K., Zheng, L., Wen, Z., Feng, Y., ... \& Chen, W. AgentCoord: Visually Exploring Coordination Strategy for LLM-based Multi-Agent Collaboration. Preprint at \url{https://arXiv.org/abs/2404.11943} (2024).

\bibitem{hong2023}
Hong, S., Zheng, X., Chen, J., Cheng, Y., Zhang, C., Wang, Z., Yau, S. K. S., Lin, Z., Zhou, L., Ran, C. \textit{et al.} Metagpt: Meta programming for multi-agent collaborative framework. Preprint at \url{https://arXiv.org/abs/2308.00352} (2023).

\bibitem{mandi2023}
Mandi, Z., Jain, S. \& Song, S. Roco: Dialectic multi-robot collaboration with large language models. Preprint at \url{https://arXiv.org/abs/2307.04738} (2023).

\bibitem{park2023}
Park, J. S., O’Brien, J. C., Cai, C. J., Morris, M. R., Liang, P. \& Bernstein, M. S. Generative agents: Interactive simulacra of human behavior. Preprint at \url{https://arXiv.org/abs/2304.03442} (2023).

\bibitem{wang2023}
Wang, S., Liu, C., Zheng, Z., Qi, S., Chen, S., Yang, Q., Zhao, A., Wang, C., Song, S. \& Huang, G. Avalon’s game of thoughts: Battle against deception through recursive contemplation. Preprint at \url{https://arXiv.org/abs/2310.01320} (2023).

\bibitem{lin2024}
Lin, Q., Hu, R., Li, H., Wu, S., Li, Y., Fang, K., ... \& Xu, L. ShapefileGPT: A Multi-Agent Large Language Model Framework for Automated Shapefile Processing. Preprint at \url{https://arXiv.org/abs/2410.12376} (2024).

\bibitem{lee2025}
Lee, C., Paramanayakam, V., Karatzas, A., Jian, Y., Fore, M., Liao, H., ... \& Stamoulis, D. Multi-Agent Geospatial Copilots for Remote Sensing Workflows. Preprint at \url{https://arXiv.org/abs/2501.16254} (2025).

\bibitem{shen2024}
Shen, Y., Song, K., Tan, X., Li, D., Lu, W. \& Zhuang, Y. Hugginggpt: Solving ai tasks with chatgpt and its friends in hugging face. \textit{Advances in Neural Information Processing Systems} 36 (2024).

\bibitem{huang2024}
Huang, J. T., Zhou, J., Jin, T., Zhou, X., Chen, Z., Wang, W., ... \& Lyu, M. R. On the resilience of multi-agent systems with malicious agents. Preprint at \url{https://arXiv.org/abs/2408.00989} (2024).

\bibitem{rew1990}
Rew, R. \& Davis, G. NetCDF: an interface for scientific data access. \textit{IEEE Computer Graphics and Applications} 10, 76--82 (1990). \url{https://doi.org/10.1109/38.56302}

\bibitem{hoyer2017}
Hoyer, S. \& Hamman, J. xarray: N-D labeled Arrays and Datasets in Python. \textit{Journal of Open Research Software} 5, 10 (2017). \url{https://doi.org/10.5334/jors.148}

\bibitem{madaan2024}
Madaan, A., Tandon, N., Gupta, P., Hallinan, S., Gao, L., Wiegreffe, S., ... \& Clark, P. Self-refine: Iterative refinement with self-feedback. \textit{Advances in Neural Information Processing Systems} 36 (2024).

\bibitem{ferraz2024}
Ferraz, T. P., Mehta, K., Lin, Y. H., Chang, H. S., Oraby, S., Liu, S., ... \& Peng, N. LLM self-correction with DeCRIM: Decompose, critique, and refine for enhanced following of instructions with multiple constraints. Preprint at \url{https://arXiv.org/abs/2410.06458} (2024).

\bibitem{eyring2016}
Eyring, V., Bony, S., Meehl, G. A., Senior, C. A., Stevens, B., Stouffer, R. J. \& Taylor, K. E. Overview of the Coupled Model Intercomparison Project Phase 6 (CMIP6) experimental design and organization. \textit{Geoscientific Model Development} 9, 1937--1958 (2016). \url{https://doi.org/10.5194/gmd-9-1937-2016}

\bibitem{liu2024}
Liu, N., Chen, L., Tian, X., Zou, W., Chen, K. \& Cui, M. From llm to conversational agent: A memory enhanced architecture with fine-tuning of large language models. Preprint at \url{https://arXiv.org/abs/2401.02777} (2024).

\bibitem{jimenez2023}
Jimenez, C. E., Yang, J., Wettig, A., Yao, S., Pei, K., Press, O. \& Narasimhan, K. Swe-bench: Can language models resolve real-world github issues?. Preprint at \url{https://arXiv.org/abs/2310.06770} (2023).

\bibitem{zhang2024}
Zhang, Y., Wang, Z., He, Z., Li, J., Mai, G., Lin, J., ... \& Yu, W. BB-GeoGPT: A framework for learning a large language model for geographic information science. \textit{Information Processing \& Management} 61, 103808 (2024).

\bibitem{gao2025}
Gao, M., Ye, Y., Zheng, Y. \& Lai, J. A comprehensive analysis of R's application in ecological research from 2008 to 2023. \textit{Journal of Plant Ecology} rtaf010 (2025). \url{https://doi.org/10.1093/jpe/rtaf010}

\bibitem{pei2024}
Pei, D., He, J., Liu, K., Chen, M. \& Zhang, S. Application of large language models and assessment of their ship-handling theory knowledge and skills for connected maritime autonomous surface ships. \textit{Mathematics} 12(15), 2381 (2024).

\bibitem{wynn2014}
Wynn, R. B. \textit{et al.} Autonomous Underwater Vehicles (AUVs): Their past, present and future contributions to the advancement of marine geoscience. \textit{Marine Geology} 352, 451--468 (2014).
\end{thebibliography}
\end{document}